\documentclass
[floatfix,superscriptaddress,secnumarabic,amssymb,amsmath,nobibnotes,aps,prd,showkeys,showpacs,nofootinbib,onecolumn,notitlepage,12pt]{revtex4}%
\usepackage{setspace}
\usepackage{color}
\usepackage{amsmath}
\usepackage{amsfonts}
\usepackage{verbatim}
\usepackage{amssymb}
\usepackage{graphicx,bm}
\usepackage{graphicx}
\usepackage{amsmath}
\usepackage{amssymb}
\usepackage{amssymb}
\usepackage{graphicx,bm}
\usepackage{graphicx}
\usepackage[caption=false]{subfig}%
\providecommand{\U}[1]{\protect\rule{.1in}{.1in}}
\newcommand{\be}{\begin{equation}}
\newcommand{\ee}{\end{equation}}

\newcommand{\mincir}{\raise
-3.truept\hbox{\rlap{\hbox{$\sim$}}\raise4.truept\hbox{$<$}\ }}
\newcommand{\magcir}{\raise
-3.truept\hbox{\rlap{\hbox{$\sim$}}\raise4.truept\hbox{$>$}\ }}

\ifx\pdfoutput\relax\let\pdfoutput=\undefined\fi
\newcount\msipdfoutput
\ifx\pdfoutput\undefined\else
\ifcase\pdfoutput\else
\ifx\paperwidth\undefined\else
\ifdim\paperheight=0pt\relax\else\pdfpageheight\paperheight\fi
\ifdim\paperwidth=0pt\relax\else\pdfpagewidth\paperwidth\fi
\fi\fi\fi
\begin{document}
\title{Dipole Cosmology in $f\left(  Q\right)  $-gravity}
\author{Andronikos Paliathanasis}
\email{anpaliat@phys.uoa.gr}
\affiliation{Institute of Systems Science, Durban University of Technology, Durban 4000,
South Africa}
\affiliation{Departamento de Matem\'{a}ticas, Universidad Cat\'{o}lica del Norte, Avda.
Angamos 0610, Casilla 1280 Antofagasta, Chile}

\begin{abstract}

\end{abstract}
\begin{abstract}
Symmetric teleparallel $f(Q)$-gravity allows for the presence of a perfect
fluid with a tilted velocity in the Kantowski-Sachs geometry. In this dipole
model, we consider an ideal gas and we investigate the evolution of the
physical parameters. The tilt parameter is constrained by the nonlinear
function $f(Q)$ through the non-diagonal equations of the field equations. We
find that the dynamics always reduce to the vacuum solutions of STEGR. This
includes the Kasner universe, when no cosmological term is introduced by the
$f(Q)$ function, and the isotropic de Sitter universe, where $f\left(
Q\right)  $ plays the role of the cosmological constant. In the extreme tilt
limit, the universe is consistently anisotropic and accelerated. However, the
final solution matches that of STEGR.

\end{abstract}
\maketitle

\section{Introduction}

\label{sec1}

The cosmological principle states that the universe is isotropic and
homogeneous. In the framework of General Relativity, the physical space is
described by the Friedmann--Lema\^{\i}tre--Robertson--Walker (FLRW) geometry,
the matter source is expressed in terms of perfect fluids an the observer
which is defined to be orthogonal to the homogeneous and isotropic surface.
The analysis of cosmological observations for the early and late-time stages
of the universe gives rise to tensions in the physical parameters
\cite{t1,t2,t3}.

Cosmological tensions have prompted the exploration of new gravitational
theories, including those that challenge the cosmological principle \cite{t4}.
The small anisotropies and inhomogeneities in the CMB \cite{pl2018} indicate
that at the early stages of the universe the cosmological principle was
violated, that has lead to the study of anisotropic and inhomogeneous
cosmological solutions \cite{in1,in2,in3}. Inhomogeneous and anisotropic
cosmologies are the models where the limit of FLRW is recovered \cite{kras}.

Bianchi cosmologies contain cosmological models that have been utilized to
discuss the anisotropies of the primordial universe and its evolution toward
the observed isotropy of the present epoch
\cite{Mis69,jacobs2,collins,JB1,JB2,jb33}. It has been shown \cite{f3} that
some homogeneous models which start from inhomogeneous models can become
anisotropic in the future. However, because of the inflation that will happen
in an exponentially distant time in the future, and in the present era the
models to be still homogeneous metric perturbations. Primordial anisotropies
have been investigated before in a various of studies \cite{pa1,pa2}. On the
other hand, there are various studies which investigate the small anisotropies
in the recent cosmological observations, see for instance
\cite{pl1,pl2,pl3,pl4}. On the other hand, the nonzero value of the spatial
curvature in the early stages of the universe is not excluded by the
observations, see the discussion in \cite{val1}.

In 1972, King and Ellis \cite{ke1} investigated the existence of homogeneous
spacetimes characterized by a perfect fluid with tilted velocity. Tilted
cosmological models are characterized by non-zero expansion, rotation, and
shear of the fluid source. These characteristics led to the discovery of new
singular solutions in Bianchi cosmologies \cite{ke2}. Bianchi II tilted
cosmological models are characterized by vanishing rotation, whereas Bianchi V
and Bianchi IV universes can exist with or without rotation. Locally
rotational spacetimes with a tilted source are specifically the Bianchi V and
Bianchi VII models \cite{ke1}. The presence of non-zero rotation in tilted
cosmological solutions can result in these solutions appearing inhomogeneous
to another observer, even if the original spacetime is spatially homogeneous
\cite{ke1}.

The Bianchi V cosmology with a tilted ideal gas was studied in detail by
Hewitt in \cite{hew}. In this work, the global dynamics of the cosmological
parameters were explored for a time-varying tilted parameter. It was
discovered that the limit of the cosmological principle exists as a future
attractor, while the Kasner spacetime can describe the asymptotic solution
near the singularity. Subsequently, a more general treatment of the dynamics
of homogeneous tilted cosmologies was presented in \cite{col0,col0a,col0b}. In
a related context, the asymptotic dynamics of an inhomogeneous tilted
cosmology was introduced in \cite{col1} for the analytic study of the Cosmic
Microwave Background (CMB). Tilted models have been employed in various
studies in the literature to explain observational data, as seen in works such
as \cite{st1,st2}. Additionally, studies on tilted cosmologies where the fluid
has a time-dependent equation of state parameter can be found in
\cite{tt1,tt2}. These works contribute to our understanding of tilted
cosmological models.

On the other hand, alternative and modified theories of gravity have been
proposed in recent decades by cosmologists to provide explanations for various
observational phenomena. These alternative and modified gravitational theories
introduce new degrees of freedom, which in turn open up new directions in the
study of astrophysical objects and the evolution of the cosmos. These theories
aim to address unresolved questions and discrepancies in our current
understanding of gravity and cosmology, offering potential avenues for
exploring the behavior of the universe on large scales. Recently, the
Symmetric Teleparallel Equivalent of General Relativity (STEGR)
\cite{Nester:1998mp} and its extensions
\cite{Koivisto2,Koivisto3,gg1,jjd1,pal2} have drawn the attention of
cosmologists. In STEGR the spacetime is defined by a metric tensor, while the
autoparallels are defined with the use of a symmetric and flat connection
different from the Levi-Civita connection. Consequently, the curvature
$R_{\;\lambda\mu\nu}^{\kappa}$ and the torsion tensors $\mathrm{T}_{\mu\nu
}^{\lambda}$ for this connection are always zero, and only the nonmetricity
tensor survives~$\nabla_{\kappa}g_{\mu\nu}=Q_{\kappa\mu\nu}\neq0~$%
\cite{Eisenhart}. In STEGR the corresponding Einstein-Hilbert Action is
defined by the nonmetricity scalar $Q$, which plays the fundamental role of
STEGR and its modifications.

$f\left(  Q\right)  $-gravity \cite{Koivisto2,Koivisto3} is one of the
simplest extensions of STEGR. The gravitational Lagrangian is defined by a
nonlinear function $f$ of the nonmetricity scalar. The theory has found
various applications in cosmological studies
\cite{ww0,ww1,ww2,ww3,ww5,ww6,ww8,ww10,ww11,ww13,ww14} and in compact objects
\cite{sw1,sw2,sw3,sw4,sw5}. For a recent review in $f\left(  Q\right)
$-gravity we refer the reader in \cite{revh}. Although $f\left(  Q\right)
$-gravity is charged for the appearance of ghosts or of strong coupling in
FLRW background \cite{ppr1,ppr2}, it provides unique directions for the study
of gravitational models. Nevertheless the introduction of matter source or
scalar fields, or the consideration of of another geometry different from that
of the FLRW background, can overpass the limits of $f\left(  Q\right)
$-gravity. See for instance the discussion in \cite{stg1} for teleparallelism.

One of the main characteristics of STEGR and its generalizations is the
ambiguity in defining the connection, leading to the existence of multiple
gravitational theories for the same gravitational model. While these theories
converge to the same predictions in the limit of STEGR, they diverge in
modified theories of gravity. Specifically, for the spatially flat FLRW
geometry,there are three families of symmetric and flat connections
\cite{Hohmann,Heis2,Zhao}, resulting in an equal number of gravitational
theories with distinct cosmological dynamics and evolution \cite{an01}. This
ambiguity in the choice of connection has implications for the analysis of
cosmological observations, as different connections can lead to different
interpretations of observational data \cite{ww11}.

In this piece of work, we focus on the locally rotationally symmetric
Kantowski-Sachs spacetime within the framework of $f\left(  Q\right)
$-gravity. Previous research has identified two families of connections that
are symmetric, flat, and invariant under the four isometries of the
Kantowski-Sachs metric \cite{ks1}. These connections result in a gravitational
theory with a modified Einstein tensor equation featuring non-diagonal
components. The non-diagonal terms can be removed through additional
constraints on the $f\left(  Q\right)  $ function or on the connection itself.

For the first connection, the non-diagonal terms vanish when $f\left(
Q\right)  $ is a linear function, leading to the recovery of STEGR. On the
other hand, the second connection eliminates the non-diagonal components by
reducing the degrees of freedom in the connection. This latter connection has
been the subject of previous studies, where exact scaling solutions were
determined in \cite{ks1}, and the asymptotic analysis of the phase-space was
recently investigated in \cite{ks2}. For further exploration of homogeneous
and anisotropic spacetimes we refer the reader in \cite{bb1,bb2} and
references therein.

In the following, we consider the first connection for the Kantowski-Sachs
geometry in $f\left(  Q\right)  $-gravity; and in order the theory to survive
for a nonlinear function $f\left(  Q\right)  $ we introduce a perfect fluid
with a tilted velocity. Recall that while in General Relativity
Kantowski-Sachs geometry does not support a tilted velocity, that is not the
case in $f\left(  Q\right)  $-gravity, that is, due to the nature of the
theory. We extend the analysis of \cite{hew,col0} for the latter gravitational
model. We perform a detailed analysis of the asymptotics for a perfect fluid
with constant equation of state parameter. We consider that the tilted
parameter is time-dependent where we investigate also the limit of extreme
tilted velocity. In General Relativity, the tilted velocity is constraint
through the field equations with the spatial curvature of the background
geometry; in our consideration the tilted velocity is constraint with the
nonlinear function $f\left(  Q\right)  $. Without loss of generality we make
use of the scalar field description of $f\left(  Q\right)  $-gravity
\cite{min1}. Phase-space analysis stands as a potent method for analytically
treating nonlinear field equations and deriving asymptotic exact solutions.
Through this analysis, we aim to gain insights into the impacts of tilted
velocity in homogeneous cosmologies, while also exploring whether the theory
provides the limit of General Relativity in the presence of a tilted observer.
The structure of the paper is as follows.

In Section \ref{sec2}, we provide the fundamental definitions of
$f(Q)$-gravity. Section \ref{sec3} is dedicated to discussing the homogeneous
and anisotropic Kantowski-Sachs geometry. Here, we utilize previous findings
to express the dynamical degrees of freedom of the $f(Q)$ function in terms of
scalar fields. Section \ref{sec4} introduces the dipole cosmological model,
where we consider a pressureless perfect fluid with the tilted velocity
oriented perpendicular to the two-dimensional sphere of the background
spacetime. Moving on to Section \ref{sec5}, we delve into the phase-space
analysis for this cosmological model. Our focus is on the power-law
$f(Q)=f_{0}Q^{\alpha}$, although our main results are generalizable to any
function $f(Q)$ that asymptotically behaves as a power-law. We particularly
highlight the extreme tilt limit in this analysis. Finally, in Section
\ref{sec6}, we summarize our findings and draw conclusions based on the
results obtained throughout the study.

\section{$f\left(  Q\right)  $-gravity}

\label{sec2}

We consider the four-dimensional manifold that describes physical space
characterized by the metric tensor $g_{\mu\nu}$ and a symmetric and flat
connection $\Gamma_{\mu\nu}^{\kappa}$ with the property%

\begin{equation}
\nabla_{\kappa}g_{\mu\nu}=Q_{\kappa\mu\nu}\equiv\frac{\partial g_{\mu\nu}%
}{\partial x^{\lambda}}-\Gamma_{\;\lambda\mu}^{\sigma}g_{\sigma\nu}%
-\Gamma_{\;\lambda\nu}^{\sigma}g_{\mu\sigma},
\end{equation}
in which $Q_{\kappa\mu\nu}$ is the cotorsion tensor, also known as the
nonmetricity tensor \cite{Eisenhart}. \newline

Because the connection $\Gamma_{\mu\nu}^{\kappa}$ is symmetric, there is not
any torsion component, that is,%
\begin{equation}
\mathrm{T}_{\mu\nu}^{\lambda}\equiv\Gamma_{\;\left[  \mu\nu\right]  }%
^{\lambda}=0,
\end{equation}
and the curvature tensor for the connection $\Gamma_{\mu\nu}^{\kappa}~$is
always zero
\begin{equation}
R_{\;\lambda\mu\nu}^{\kappa}\equiv\frac{\partial\Gamma_{\;\lambda\nu}^{\kappa
}}{\partial x^{\mu}}-\frac{\partial\Gamma_{\;\lambda\mu}^{\kappa}}{\partial
x^{\nu}}+\Gamma_{\;\lambda\nu}^{\sigma}\Gamma_{\;\mu\sigma}^{\kappa}%
-\Gamma_{\;\lambda\mu}^{\sigma}\Gamma_{\;\mu\sigma}^{\kappa}=0.
\end{equation}

The gravitational Action of $f\left(  Q\right)  $-gravity is defined as
\cite{Koivisto2,Koivisto3}
\begin{equation}
S_{f\left(  Q\right)  }=\int d^{4}x\sqrt{-g}f(Q),
\end{equation}
where $Q$ is the nonmetricity scalar defined as \cite{Nester:1998mp}%
\[
Q=Q_{\lambda\mu\nu}P^{\lambda\mu\nu},
\]
tensor $P^{\lambda\mu\nu}$ is defined as
\begin{equation}
P_{\;\mu\nu}^{\lambda}=-\frac{1}{4}Q_{\;\mu\nu}^{\lambda}+\frac{1}{2}%
Q_{(\mu\phantom{\lambda}\nu)}^{\phantom{(\mu}\lambda\phantom{\nu)}}+\frac
{1}{4}\left(  Q^{\lambda}-\bar{Q}^{\lambda}\right)  g_{\mu\nu}-\frac{1}%
{4}\delta_{\;(\mu}^{\lambda}Q_{\nu)},
\end{equation}
and $Q_{\mu}=Q_{\mu\nu}^{\phantom{\mu\nu}\nu}$ and $\bar{Q}_{\mu
}=Q_{\phantom{\nu}\mu\nu}^{\nu\phantom{\mu}\phantom{\mu}}$.

If $\overset{o}{R}$ describes the Ricci scalar defined by the Levi-Civita
connection for the metric tensor $g_{\mu\nu}$, it holds that \cite{revh}
\begin{equation}
\int d^{4}x\sqrt{-g}Q\simeq\int d^{4}x\sqrt{-g}\overset{o}{R}+\text{boundary
terms.}%
\end{equation}
This property implies that in the linear case, $f(Q)$-gravity is equivalent to
General Relativity. This theory is commonly referred to as Symmetric
Teleparallel General Relativity (STGR).

Varying of the Action Integral $S_{f\left(  Q\right)  }$ with respect to the
metric tensor gives the modified field equations \cite{revh}
\begin{equation}
\frac{2}{\sqrt{-g}}\nabla_{\lambda}\left(  \sqrt{-g}f^{\prime}(Q)P_{\;\mu\nu
}^{\lambda}\right)  -\frac{1}{2}f(Q)g_{\mu\nu}+f^{\prime}(Q)\left(  P_{\mu
\rho\sigma}Q_{\nu}^{\;\rho\sigma}-2Q_{\rho\sigma\mu}P_{\phantom{\rho\sigma}\nu
}^{\rho\sigma}\right)  =0, \label{fe.01}%
\end{equation}
where $f^{\prime}\left(  Q\right)  =\frac{df}{dQ}$.

On the other hand, variation with respect to the connection gives the equation
of motion \cite{revh}
\begin{equation}
\nabla_{\mu}\nabla_{\nu}\left(  \sqrt{-g}f^{\prime}%
(Q)P_{\phantom{\mu\nu}\sigma}^{\mu\nu}\right)  =0. \label{fe.02}%
\end{equation}

The equation of motion for the connection is possible be identically
satisfied, in this case we shall say that the connection is defined in the
\textquotedblleft coincidence gauge\textquotedblright.

It is important to note that the connection and the metric tensor have
different transformation rules, and the connection is coordinate-dependent.
While the flatness of the connection implies that all its components can be
zero in a particular coordinate system, the choice of a specific metric tensor
already defines the coordinate system. In this specific coordinate system, if
the equation of motion for the connection is not trivially zero, we say that
the connection is defined in the \textquotedblleft non-coincidence
gauge\textquotedblright. On the other hand, it has been shown that in the
presence of spatial curvature the limit of General Relativity is recovered for
connections defined in the non-coincidence gauge \cite{Zhao}.

\section{Kantowski-Sachs geometry}

\label{sec3}

We introduce the Kantowski-Sachs spacetime expressed in the Misner variables
where the line element is given by
\begin{equation}
ds^{2}=-dt^{2}+a^{2}\left(  t\right)  \left(  e^{2b\left(  t\right)  }%
dr^{2}+e^{-b\left(  t\right)  }\left(  d\theta^{2}+\sin^{2}\theta d\varphi
^{2}\right)  \right)  . \label{ks.01}%
\end{equation}
Function $a\left(  t\right)  $ is the scale factor which defines the radius of
the space and function $b\left(  t\right)  $ defines the anisotropy.
Isotropization is recovered when $b\left(  t\right)  =0$. In this limit the
Kantowski-Sachs spacetime describes the closed FLRW geometry. For the comoving
observe, the Hubble function is defined as $H=\frac{\dot{a}}{a}$.

Kantowski-Sachs universe \cite{ks1p} is described by an homogeneous and
anisotropic line element with a topology $R\times S^{2}$ and admits four
Killing vector fields which act on spacelike hypersurfaces \cite{WE}. In the
isotropization limit the Kantowski-Sachs universe is reduced to that of the
closed FLRW universe. There are a plethora of applications in the literature
where the\ Kantowski-Sachs space has been used to explain the physical
phenomena \cite{wa1,wa2,wa3,wa5,wa4,wa7,wa8}. Another important characteristic
of Kantowski-Sachs universe is that it can be the limit of inhomogeneous and
anisotropic geometries \cite{kras}. Kantowski-Sachs geometry has some
important characteristics, indeed there can be barrel, pancake, cigar
singularities or isotropic structure based on the initial conditions
\cite{col}. That properties makes Kantowski-Sachs an important mathematical
structure for the connection of different eras in the cosmic evolution. For a
fruitful discussion on the matter we refer the reader to \cite{col10}.

The line element (\ref{ks.01}) admits four isometries, they are%
\begin{equation}
\xi_{1}=\partial_{\varphi}~,~\xi_{2}=\cos\varphi\,\partial_{\theta}-\cot
\theta\sin\varphi\partial_{\varphi}~,
\end{equation}%
\begin{equation}
\xi_{3}=\sin\varphi\partial_{\theta}+\cot\theta\cos\varphi\partial_{\varphi
}~,~\xi_{4}=\partial_{x}~.
\end{equation}

The requirement the connection $\Gamma_{\mu\nu}^{\kappa}$ to be flat, and to
inherit the isometries of the Kantowski-Sachs geometry leads to two families
of connections \cite{ks1}. These two connections are defined in the
non-coincidence gauge, where dynamical degrees of freedom are introduced by
the connection in the field equations.

In the following we employ the connection with non-zero coefficients%
\begin{equation}%
\begin{split}
&  \Gamma_{\;tt}^{t}=\gamma_{2},\quad\Gamma_{\;tt}^{r}=\frac{1}{c_{1}}\left(
\dot{\gamma}_{1}-\gamma_{1}\gamma_{2}+\gamma_{1}^{2}\right)  ,\quad
\Gamma_{\;tr}^{r}=\Gamma_{\;t\theta}^{\theta}=\Gamma_{\;tz}^{z}=\gamma
_{1},\quad\Gamma_{\;\theta\theta}^{r}=-\frac{1}{c_{1}},\\
&  \Gamma_{\;rr}^{r}=\Gamma_{\;r\theta}^{\theta}=\Gamma_{r\varphi\;}^{\varphi
}=c_{1},\quad\Gamma_{\;\varphi\varphi}^{r}=-\frac{\sin^{2}\theta}{c_{1}}%
,\quad\Gamma_{\;\varphi\varphi}^{\theta}=-\cos\theta\sin\theta,\quad
\Gamma_{\;\theta\varphi}^{\varphi}=\cot\theta.
\end{split}
\end{equation}
where $c_{1}$ is a nonzero constant and $\gamma_{1}\left(  t\right)
,~\gamma_{2}\left(  t\right)  $ are functions.

The non-metricity scalar for the latter connection and the metric
(\ref{ks.01}) is expressed as follows%
\begin{equation}
Q=-6H^{2}+\frac{3}{2}\dot{b}^{2}+2\frac{e^{b}}{a^{2}}+3\left(  3H\gamma
_{1}+\dot{\gamma}_{1}\right)  , \label{eq.00}%
\end{equation}

\subsection{Vacuum field equations}

The\ nonzero components of the gravitational field equations (\ref{fe.01}) in
the vacuum are

$tt:$%
\begin{equation}
f^{\prime}(Q)\left(  3H^{2}+\frac{k}{a^{2}}e^{b}-\frac{3}{4}\dot{b}%
^{2}\right)  +\frac{1}{2}\left(  f(Q)-Qf^{\prime}(Q)\right)  +\frac{3}%
{2}\gamma_{1}\dot{Q}f^{\prime\prime}(Q)=0. \label{eq.01}%
\end{equation}

$tr:$%
\begin{equation}
\frac{3}{2}c_{1}\dot{Q}f^{\prime\prime}=0. \label{eq.02}%
\end{equation}

$rr:$%
\begin{equation}
f^{\prime}\left(  Q\right)  \left(  \ddot{b}+3H\dot{b}-2\dot{H}-3H^{2}%
-\frac{e^{b}}{a^{2}}-\frac{3}{4}\dot{b}^{2}\right)  -\frac{1}{2}\left(
f\left(  Q\right)  -Qf^{\prime}\left(  Q\right)  \right)  +\dot{Q}%
f^{\prime\prime}\left(  Q\right)  \left(  \dot{b}-2H+\frac{3}{2}\gamma
_{1}\right)  =0.
\end{equation}

$\theta\theta,\varphi\varphi:$
\begin{equation}
f^{\prime}\left(  Q\right)  \left(  \ddot{b}+3H\acute{b}+4\dot{H}+6H^{2}%
+\frac{3}{2}\dot{b}^{2}\right)  +\left(  f\left(  Q\right)  -Qf^{\prime
}\left(  Q\right)  \right)  +\dot{Q}f^{\prime\prime}\left(  Q\right)  \left(
4H+\dot{b}-3\gamma_{1}\right)  =0.
\end{equation}

Furthermore, the equation of motion (\ref{fe.02}) for the connection takes the
following expression%
\begin{equation}
\left(  a^{3}f^{\prime\prime}\left(  Q\right)  \dot{Q}\right)  ^{\cdot}=0,
\label{eq.05a}%
\end{equation}
which is a conservation law.

We observe the nondiagonal equation (\ref{eq.02}) gives the constraint
$\dot{Q}f^{\prime\prime}=0$, that is, $f\left(  Q\right)  =f_{1}Q+f_{2}$ or
$Q=Q_{0}$. In these two cases the cosmological solution is that of General
Relativity with (or without) the cosmological constant term.

For this specific connection, the only allowed solution in vacuum corresponds
to that of General Relativity. However, due to the nature of the nonmetricity
theory, it is possible to have fluids with a tilted observer in the context of
the Kantowski-Sachs geometry. This is the scenario we investigate in the
following discussion.

At this point we remark that an equivalent way to write the field equations is
by using a scalar field description \cite{min1}.

We introduce $\phi=f^{\prime}\left(  Q\right)  $ and $V\left(  \phi\right)
=\left(  f(Q)-Qf^{\prime}(Q)\right)  ~$, then the field equations read

$tt:$%
\begin{equation}
\phi\left(  3H^{2}+\frac{k}{a^{2}}e^{b}-\frac{3}{4}\dot{b}^{2}\right)
+\frac{1}{2}V\left(  \phi\right)  +\frac{3}{2}\gamma_{1}\dot{\phi}=0,
\end{equation}

$tr:$%
\begin{equation}
\frac{3}{2}c_{1}\dot{\phi}=0.
\end{equation}

$rr:$%
\begin{equation}
\phi\left(  \ddot{b}+3H\dot{b}-2\dot{H}-3H^{2}-\frac{e^{b}}{a^{2}}-\frac{3}%
{4}\dot{b}^{2}\right)  -\frac{1}{2}V\left(  \phi\right)  +\dot{\phi}\left(
\dot{b}-2H+\frac{3}{2}\gamma_{1}\right)  =0,
\end{equation}

$\theta\theta,\varphi\varphi:$
\begin{equation}
\phi\left(  \ddot{b}+3H\dot{b}+4\dot{H}+6H^{2}+\frac{3}{2}\dot{b}^{2}\right)
+\frac{1}{2}V\left(  \phi\right)  +\dot{\phi}\left(  4H+\dot{b}-3\gamma
_{1}\right)  =0.
\end{equation}

and the equation of motion for the connection%
\begin{equation}
\left(  a^{3}\dot{\phi}\right)  ^{\cdot}=0.
\end{equation}

\section{Dipole Cosmology}

\label{sec4}

We introduce the perfect fluid energy-momentum tensor \cite{ke1,ke2}%
\begin{equation}
T_{\mu\nu}=\left(  \rho+p\right)  \tilde{u}_{\mu}\tilde{u}_{\nu}+pg_{\mu\nu},
\label{tl.02}%
\end{equation}
in which $\tilde{u}^{\mu}$ is the tilted observer, that is, $\tilde{u}^{\mu
}\tilde{u}_{\mu}=-1$, and $\tilde{u}^{\mu}=\cosh\beta\left(  t\right)
\partial_{t}+\frac{e^{-b}}{a}\sinh\beta\left(  t\right)  \partial_{r}$.
Function $\beta\left(  t\right)  $ is the tilted parameter, when $\beta\left(
t\right)  =0$, $\tilde{u}^{\mu}$ is reduced to the comoving observer $u^{\mu
}=\partial_{t}$.

For this observer, the energy momentum tensor (\ref{tl.02}) reads
\begin{equation}
T_{\mu\nu}=%
\begin{pmatrix}
\cosh^{2}\beta\left(  \rho+p\right)  -p & -ae^{b}\sinh\beta\cosh\beta\left(
\rho+p\right)  & 0 & 0\\
-ae^{b}\sinh\beta\cosh\beta\left(  \rho+p\right)  & a^{2}e^{2b}\left(
\cosh^{2}\beta\left(  \rho+p\right)  -\rho\right)  & 0 & 0\\
0 & 0 & a^{2}e^{-b}p & 0\\
0 & 0 & 0 & a^{2}e^{-b}p\sin^{2}\theta
\end{pmatrix}
.
\end{equation}

We assume that the matter source is minimally coupled to gravity, thus, we
derive the conservation equations%
\begin{align}
\dot{\rho}+3\left(  1+\frac{p}{\rho}\right)  H\rho+\left(  1+\frac{p}{\rho
}-2\frac{p}{\rho}\coth^{2}\beta\right)  \tanh\beta~\dot{\beta}\rho &
=0,\label{te.00a}\\
\dot{p}+\left(  1+\frac{p}{\rho}\right)  \rho\left(  H+\dot{b}+\coth\beta
~\dot{\beta}\right)   &  =0. \label{te.00b}%
\end{align}
From equation (\ref{te.00b}) it is clear that the evolution of the tilted
parameters $\beta\left(  t\right)  $ is related to the anisotropic scale
factor $b\left(  t\right)  $.

For the tilted observer we define the expansion rate \cite{ke1}
\begin{equation}
\tilde{\theta}=3\cosh\beta~H+\sinh\beta~\dot{\beta},
\end{equation}
the shear%
\begin{equation}
\tilde{\sigma}^{2}=\left(  \sqrt{\frac{3}{2}}\cosh\beta~\dot{b}+\sqrt{\frac
{2}{3}}\sinh\beta~\dot{\beta}\right)  ^{2}, \label{dc.01}%
\end{equation}
and the deceleration parameter \cite{ke1} $\tilde{q}=-1-\frac{3}{\tilde
{\theta}^{2}}\tilde{u}^{\mu}\theta_{;\mu}$, as%
\begin{equation}
\tilde{q}=-1-\frac{\left(  \cosh\beta~H+\frac{1}{3}\sinh\beta~\dot{\beta
}\right)  ^{\cdot}}{\left(  \cosh\beta~H+\frac{1}{3}\sinh\beta~\dot{\beta
}\right)  ^{2}}. \label{dc.02}%
\end{equation}

Finally, the field equations for the tilted observer reads
\begin{align}
\phi\left(  3H^{2}+\frac{k}{a^{2}}e^{b}-\frac{3}{4}\dot{b}^{2}\right)
+\frac{1}{2}V\left(  \phi\right)  +\frac{3}{2}\gamma_{1}\dot{\phi}  &
=\cosh^{2}\beta\left(  1+w_{m}\right)  \rho-w_{m}\rho~,\label{te.01}\\
\frac{3}{2}c_{1}\dot{\phi}  &  =-ae^{b}\sinh\beta\cosh\beta\left(
1+w_{m}\right)  \rho~,\label{te.02}\\
\phi\left(  -2\dot{H}-3H^{2}-\frac{3}{4}\dot{b}^{2}-\frac{e^{b}}{3a^{3}%
}\right)  -2\dot{\phi}H+\frac{3}{2}\dot{\phi}\gamma_{1}-V\left(  \phi\right)
&  =\left(  w_{m}+\frac{1}{3}\left(  1+w_{m}\right)  \sinh^{2}\beta\right)
\rho~,\label{te.03}\\
\phi\left(  \ddot{b}+3H\dot{b}-\frac{2}{3}\frac{e^{b}}{a^{2}}\right)  -\dot
{b}\dot{\phi}  &  =\frac{2}{3}\left(  1+w_{m}\right)  \sinh^{2}\beta
\rho~,\label{te.04}\\
\left(  a^{3}\phi\right)  ^{\cdot}  &  =0~, \label{te.05}%
\end{align}
where $w_{m}$ is the equation of state parameter for the perfect fluid, that
is, $p=w_{m}\rho$.

Equation (\ref{te.02}) states that $\dot{\phi}\neq0,~$i.e. $\dot{Q}%
f^{\prime\prime}\neq0,$ when there is a nonzero contribution of the\ tilted
fluid source in the universe. Only in this case the $f\left(  Q\right)
$-theory introduce an \textquotedblleft exotic\textquotedblright\ matter
source in the field equations.

\section{Asymptotic solutions}

\label{sec5} In this section, we conduct a comprehensive analysis of the
phase-space for the cosmological field equations (\ref{te.01})-(\ref{te.05})
considering a tilted observer with the energy-momentum tensor (\ref{tl.02}).

For the perfect fluid, we assume it to be pressureless, i.e., $p=0,~w_{m}=0$,
representing the dark matter component of the universe. Consequently,
equations (\ref{te.00a}) and (\ref{te.00b}) simplify as follows%

\begin{align}
\dot{\rho}+\left(  3H+\tanh\beta~\dot{\beta}\right)  \rho &  =0,\\
\rho\left(  H+\dot{b}+\coth\beta~\dot{\beta}\right)   &  =0.
\end{align}
In the limit where $\beta=0$, then we end with the cosmological model of STEGR
with a matter source, where the $\Lambda$CDM model is recovered in the
isotropic limit.

We introduce the new dependent dimensionless variables
\begin{equation}
\Sigma=\frac{\dot{b}}{2D}~,~x=\frac{3\dot{\phi}}{2\phi D}~,~y=\frac{V\left(
\phi\right)  }{\phi D^{2}}~,~z=\frac{\gamma}{D}~,~\mu=\sinh\beta~,~\eta
=\frac{H}{D}%
\end{equation}%
\begin{equation}
\omega_{m}=\frac{\left(  1+\cosh\left(  2\beta\right)  \right)  }{6\phi}%
\frac{\rho}{D^{2}}~,~\lambda=\phi\frac{V_{,\phi}}{V}~,~u=\left(  \frac{a}%
{D}\right)  ^{3}~,~D=\sqrt{H^{2}+\frac{e^{b}}{3a^{2}}},
\end{equation}
and the independent variable $dt=Dd\tau$.

In terms of the new variables the field equations takes the form of the
following system
\begin{align}
\frac{d\Sigma}{d\tau}  &  =\left(  \left(  \eta^{2}-1\right)  \left(
\frac{2\sqrt{3}}{3}x\Sigma+\Sigma^{2}-1\right)  +\frac{3}{2}\eta\Sigma\left(
y-xz+\Sigma^{2}-1\right)  +\frac{\mu^{2}\omega_{m}\left(  2+\eta\Sigma\right)
}{2\left(  1+\mu^{2}\right)  }\right)  ~,\label{dn.01}\\
\frac{dx}{d\tau}  &  =x\left(  \left(  \eta^{2}-1\right)  \left(  \frac
{2\sqrt{3}}{3}x+\Sigma\right)  +\frac{1}{2}\eta\left(  3\left(  y-xz+\Sigma
^{2}-1\right)  +\frac{\mu^{2}\omega_{m}}{1+\mu^{2}}\right)  \right)
~,\label{dn.02}\\
\frac{dy}{d\tau}  &  =\frac{1}{3}y\left(  x\left(  2\sqrt{3}\left(
\lambda-1+2\eta^{2}\right)  -9\eta z\right)  +6\left(  \eta^{2}-1\right)
\Sigma+3\eta\left(  3\left(  1+y+\Sigma^{2}\right)  +\frac{\mu^{2}\omega_{m}%
}{1+\mu^{2}}\right)  \right)  ~,\label{dn.03}\\
\frac{dz}{d\tau}  &  =\frac{3}{2}\eta z\left(  y-xz\right)  +\frac{\sqrt{3}%
}{3}\left(  4\eta^{2}+\left(  \lambda y-2\left(  1+\Sigma^{2}\right)  \right)
\right) \nonumber\\
&  +z\left(  \frac{\eta}{3}\left(  \eta\left(  2\sqrt{3}x+3\Sigma\right)
+\frac{9}{2}\left(  \Sigma^{2}-1\right)  +\frac{\mu^{2}\omega_{m}}{1+\mu^{2}%
}\right)  -\Sigma\right)  ~,\label{dn.04}\\
\frac{d\mu}{d\tau}  &  =-\mu\left(  \eta+2\Sigma\right)  ~,~\label{dn.05}\\
\frac{d\eta}{d\tau}  &  =\frac{1}{2}\left(  \eta^{2}-1\right)  \left(
1+3y+\frac{x}{3}\left(  4\sqrt{3}\eta-z\right)  +2\eta\Sigma+3\Sigma^{2}%
+\frac{\mu^{2}\omega_{m}}{1+\mu^{2}}\right)  ~. \label{dn.06}%
\end{align}

Furthermore, from equations (\ref{te.01}) and (\ref{te.02}) we determine the
algebraic constraints%
\begin{align}
\omega_{m}-1+y+xz-\Sigma^{2}~  &  =0,\\
\frac{\sqrt{3}}{9}c_{1}\left(  1+\mu^{2}\right)  x-u\left(  \eta^{2}-1\right)
\mu\omega_{m}  &  =0.
\end{align}
As far as the parameter $\lambda$ is concerned, we consider to be always a
constant, which corresponds to the power-law potential $V\left(  \phi\right)
$. We remark that a power-law potential corresponds to a power-law $f\left(
Q\right)  $ function. Although this is a special function $f\left(  Q\right)
$ our analysis stands in the limit when the power-law term of a given function
$f\left(  Q\right)  $ dominates the cosmological dynamics.

The canonical anisotropic parameter $\tilde{\Sigma}=\left(  \frac
{\tilde{\sigma}}{\tilde{\theta}}\right)  $ and the deceleration parameter
$\tilde{q}$ are defined for the tilted observer by expressions (\ref{dc.01})
and (\ref{dc.02}) in terms of the dimensionless variables become%
\begin{equation}
\tilde{\Sigma}^{2}=\left(  \frac{3\Sigma+\mu^{2}\left(  \Sigma-\eta\right)
}{\eta\left(  3+2\mu^{2}\right)  -2\mu^{2}\Sigma}\right)  ^{2},
\end{equation}%
\begin{align}
\frac{\tilde{q}+1}{\Delta}  &  =9y\left(  3+2\mu^{2}\left(  3+2\mu^{2}\right)
\right)  +3\left(  3\left(  1+2\mu^{2}\right)  \right)  ^{2}+\eta^{2}\left(
6+4\mu^{2}\left(  2+\mu^{2}\right)  \right) \nonumber\\
&  -8\eta\mu^{2}\left(  2+\mu^{2}\right)  \Sigma+\left(  9-4\mu^{2}\left(
1+2\mu^{2}\right)  \Sigma^{2}\right) \nonumber\\
&  +x\left(  4\sqrt{3}\left(  1+\mu^{2}\right)  \left(  \eta\left(  3+2\mu
^{2}\right)  -2\mu^{2}\Sigma\right)  -9z\left(  3+4\mu^{2}\right)  \right)  .
\end{align}
where $\Delta=\left(  6\sqrt{1+\mu^{2}}\left(  \eta\left(  3+2\mu^{2}\right)
-2\mu^{2}\Sigma^{2}\right)  ^{2}\right)  $.

\subsection{Stationary points}

We compute the stationary points of the dynamical system (\ref{dn.01}%
)-(\ref{dn.06}). Each stationary point corresponds to an asymptotic solution
for the cosmological model under study. To reconstruct the cosmological
history predicted by the model, we derive the cosmological parameters at these
stationary points. Additionally, we determine the stability properties of
these points.

The stationary points $\left(  \Sigma\left(  P\right)  ,x\left(  P\right)
,y\left(  P\right)  ,z\left(  P\right)  ,\mu\left(  P\right)  ,\eta\left(
P\right)  \right)  ~$are

\begin{itemize}
\item
\[
T_{1}=\left(  -1,0,0,\frac{4}{\sqrt{3}},\mu_{1},2\right)
\]
describes an asymptotic solution with physical parameters $\omega_{m}\left(
T_{1}\right)  =0$,~$\tilde{\Sigma}^{2}\left(  T_{1}\right)  =\frac{1}{4}$ and
$\tilde{q}\left(  T_{1}\right)  =-1+\frac{1}{2\sqrt{1+\mu_{1}^{2}}}$. Hence,
the asymptotic solution describes an accelerated Kantowski-Sachs universe. The
eigenvalues of the linearized system around the stationary point are $\left\{
-6,3,3,3,0,0\right\}  $, from where we infer that $T_{1}$ is a saddle point.

\item
\[
T_{2}=\left(  1,0,0,-\frac{4}{\sqrt{3}},\mu_{2},-2\right)  ,
\]
describes an asymptotic solution with the same physical properties of point
$T_{1}$; that is, $\omega_{m}\left(  T_{2}\right)  =0$,~$\tilde{\Sigma}%
^{2}\left(  T_{2}\right)  =\frac{1}{4}$ and $\tilde{q}\left(  T_{2}\right)
=-1+\frac{1}{2\sqrt{1+\mu_{2}^{2}}}$. The eigenvalues of the linearized system
are $\left\{  6,-3,-3,-3,0,0\right\}  $, from where we infer that $T_{2}$ is
always a saddle point.
\end{itemize}

These two points are the unique stationary points with a nonzero tilted
parameters. The following points have $\mu\left(  \Sigma\right)  =0$.

\begin{figure}[ptb]
\centering\includegraphics[width=1\textwidth]{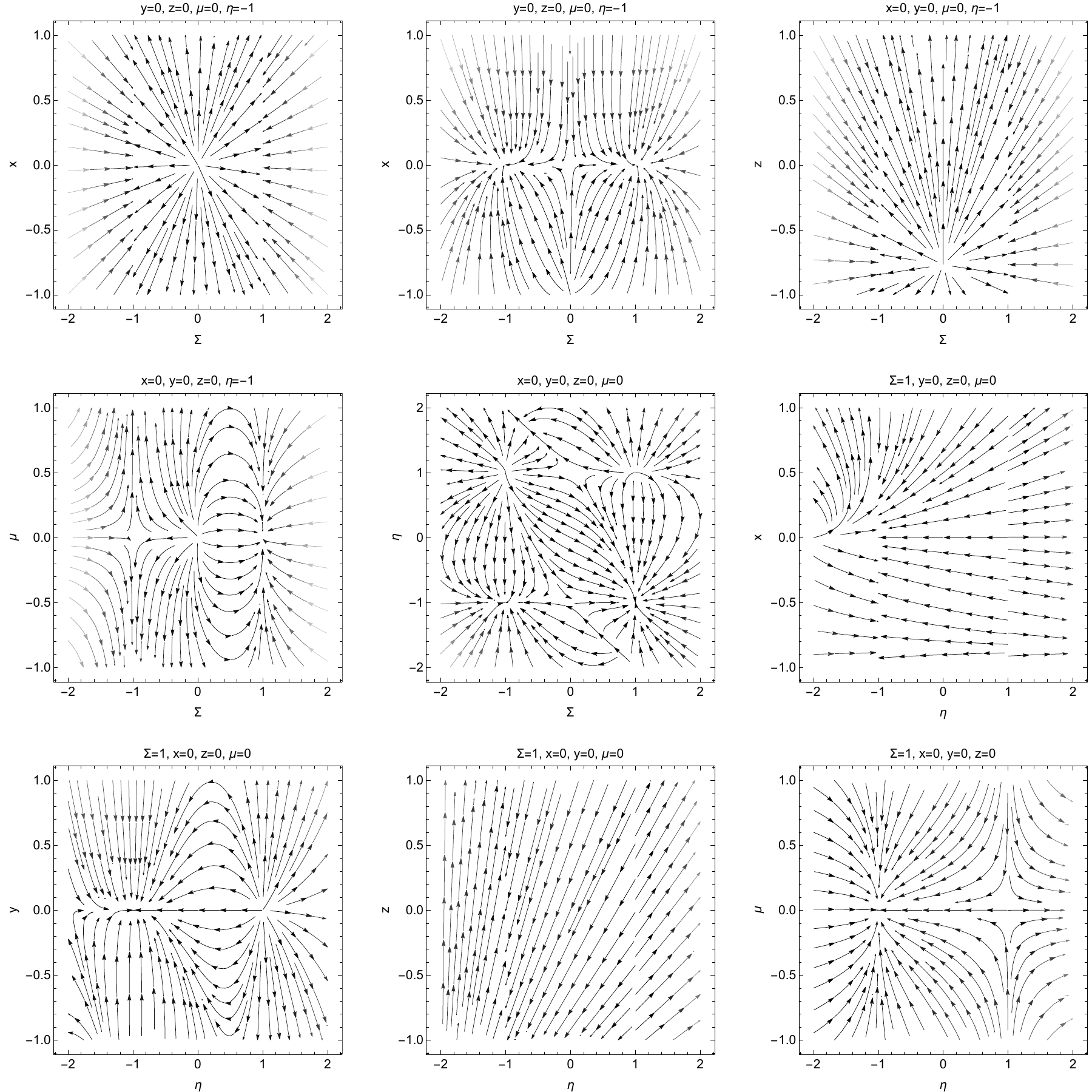}\caption{2D
Phase-space portraits for the dynamical system (\ref{dn.01})-(\ref{dn.06}) on
the surfaces where point $B_{3}$ exists. The plots are for $\lambda=1$. It
follows that $B_{3}$ is an attractor.}%
\label{fig1}%
\end{figure}

\begin{figure}[ptb]
\centering\includegraphics[width=1\textwidth]{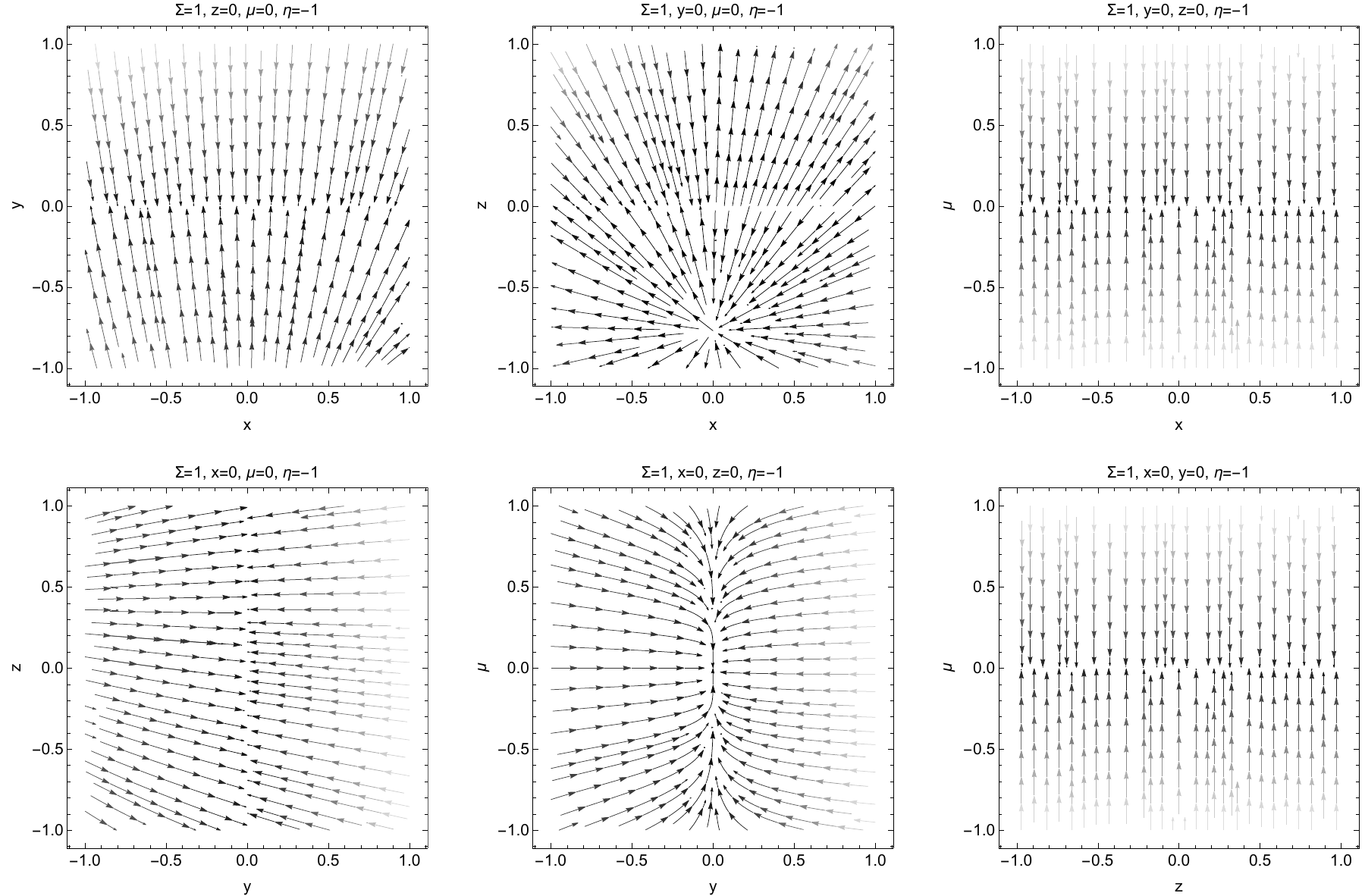}\caption{2D
Phase-space portraits for the dynamical system (\ref{dn.01})-(\ref{dn.06}) on
the surfaces where point $B_{3}$ exist. The plots are for $\lambda=1$. It
follows that $B_{3}$ is an attractor.}%
\label{fig2}%
\end{figure}

\begin{itemize}
\item
\[
B_{1}=\left(  -1,0,0,z_{1},0,-1\right)  ,
\]
with physical parameters $\omega_{m}\left(  B_{1}\right)  =0$,~$\tilde{\Sigma
}^{2}\left(  B_{1}\right)  =1$ and $\tilde{q}\left(  B_{1}\right)  =0$,
describes the Bianchi I spacetime, in particular the vacuum Kasner solution
given by General Relativity. We calculate the eigenvalues $\left\{
-6,-6,-3,3,0,0\right\}  $. Hence, $B_{1}$ is a saddle point.

\item
\[
B_{2}=\left(  -1,0,0,z_{2},0,1\right)  ,
\]
with $\omega_{m}\left(  B_{2}\right)  =0$,~$\tilde{\Sigma}^{2}\left(
B_{2}\right)  =1$, $\tilde{q}\left(  B_{2}\right)  =0$ and eigenvalues
$\left\{  6,3,2,1,0,0\right\}  $. Point $B_{2}$ describes an unstable Kasner
solution. Specifically $B_{2}$ is a source point.

\item
\[
B_{3}=\left(  1,0,0,z_{3},0,-1\right)  ,
\]
with $\omega_{m}\left(  B_{3}\right)  =0$,~$\tilde{\Sigma}^{2}\left(
B_{3}\right)  =1$, $\tilde{q}\left(  B_{3}\right)  =0$ and eigenvalues
$\left\{  -6,-3,-2,-1,0,0\right\}  $. The asymptotic solution describes the
Kasner spacetime of General Relativity.\ Because of the zero eigenvalues we
study the stability properties of point $B_{3}~$numerically. In Figs.
\ref{fig1} and \ref{fig2} we present the two-dimensional phase-space portraits
where point $B_{3}$ lies. We observe that for initial conditions where
$z\left(  B_{3}\right)  \leq0$ and$~x\left(  B_{3}\right)  >0$, point $B_{3}$
is an attractor.

\item
\[
B_{4}=\left(  1,0,0,z_{3},0,1\right)  ,
\]
describes an asymptotic solution with physical parameters $\omega_{m}\left(
B_{4}\right)  =0$,~$\tilde{\Sigma}^{2}\left(  B_{4}\right)  =1$, $\tilde
{q}\left(  B_{4}\right)  =0$ while the eigenvalues of the linearized system
around the point are $\left\{  6,6,3,-3,0,0\right\}  $. Hence, $B_{4}$ is a
saddle point which provides the Kasner solution of STEGR.

\item
\[
K_{1}=\left(  \frac{1}{2},0,-\frac{3}{4},-\frac{2+\lambda}{2\sqrt{3}}%
,0,\frac{1}{2}\right)  \text{,}%
\]
corresponds to the asymptotic solution of Kantowski-Sachs spacetime with a
cosmological constant of General Relativity. Indeed, the physical parameters
are $\omega_{m}\left(  K_{1}\right)  =0$, $\tilde{\Sigma}^{2}\left(
K_{1}\right)  =1$, $\tilde{q}\left(  K_{1}\right)  =-1$. Furthermore, the
eigenvalues of the linearized system around the point are $\left\{
-3,-\frac{3}{2},-\frac{3}{2},-\frac{3}{2},-\frac{3}{2},\frac{3}{2}\right\}  $,
which means that $K_{1}$ is a saddle point.

\item
\[
K_{2}=\left(  -\frac{1}{2},0,-\frac{3}{4},\frac{2+\lambda}{2\sqrt{3}}%
,0,-\frac{1}{2}\right)  ,
\]
describes an accelerated Kantowski-Sachs spacetime with the same physical
properties of point $K_{1}$, that is, $\omega_{m}\left(  K_{2}\right)  =0$,
$\tilde{\Sigma}^{2}\left(  K_{2}\right)  =1$, $\tilde{q}\left(  K_{2}\right)
=-1$ and eigenvalues $\left\{  3,\frac{3}{2},\frac{3}{2},\frac{3}{2},\frac
{3}{2},-\frac{3}{2}\right\}  $. Thus, $K_{2}$ is also a saddle point.

\item
\[
F_{1}=\left(  0,0,0,\frac{4}{3\sqrt{3}},0,1\right)  ,
\]
with $\omega_{m}\left(  F_{1}\right)  =1$, $\tilde{\Sigma}^{2}\left(
F_{1}\right)  =0$, $\tilde{q}\left(  F_{1}\right)  =0$, and eigenvalues
$\left\{  3,1,-1,-\frac{3}{2},-\frac{3}{2},-\frac{3}{2}\right\}  $. The
asymptotic solution is that of spatially flat FLRW universe dominated by the
matter source. Moreover, from the eigenvalues we conclude that $F_{1}$ is
always a saddle point.

\item
\[
F_{2}=\left(  0,0,0,-\frac{4}{3\sqrt{3}},0,-1\right)  ,
\]
has the same physical quantities with point $F_{1}$, that is, $\omega
_{m}\left(  F_{2}\right)  =1$, $\tilde{\Sigma}^{2}\left(  F_{2}\right)  =0$,
$\tilde{q}\left(  F_{2}\right)  =0$. The eigenvalues are calculated $\left\{
-3,-1,1,\frac{3}{2},\frac{3}{2},\frac{3}{2}\right\}  $, that is, $F_{2}$ is
always a saddle point.

\item
\[
F_{3}=\left(  0,0,-1,\frac{\lambda-2}{2\sqrt{3}},0,-1\right)  ,
\]
and physical parameters $\omega_{m}\left(  F_{3}\right)  =0$, $\tilde{\Sigma
}^{2}\left(  F_{3}\right)  =0$, $\tilde{q}\left(  F_{3}\right)  =-1$, describe
a FLRW universe with a cosmological constant. The eigenvalues are $\left\{
3,3,3,3,2,1\right\}  $, which means that that $F_{3}$ is always a source.

\item
\[
F_{4}=\left(  0,0,-1,-\frac{\lambda-2}{2\sqrt{3}},0,1\right)  ,
\]
describes the de Sitter universe, with $\omega_{m}\left(  F_{4}\right)  =0$,
$\tilde{\Sigma}^{2}\left(  F_{4}\right)  =0$, $\tilde{q}\left(  F_{4}\right)
=-1$ and eigenvalues $\left\{  -3,-3,-3,-3,-2,-1\right\}  $. Point $F_{4}$ is
the unique attractor global of the cosmological model.
\end{itemize}

\begin{figure}[ptb]
\centering\includegraphics[width=1\textwidth]{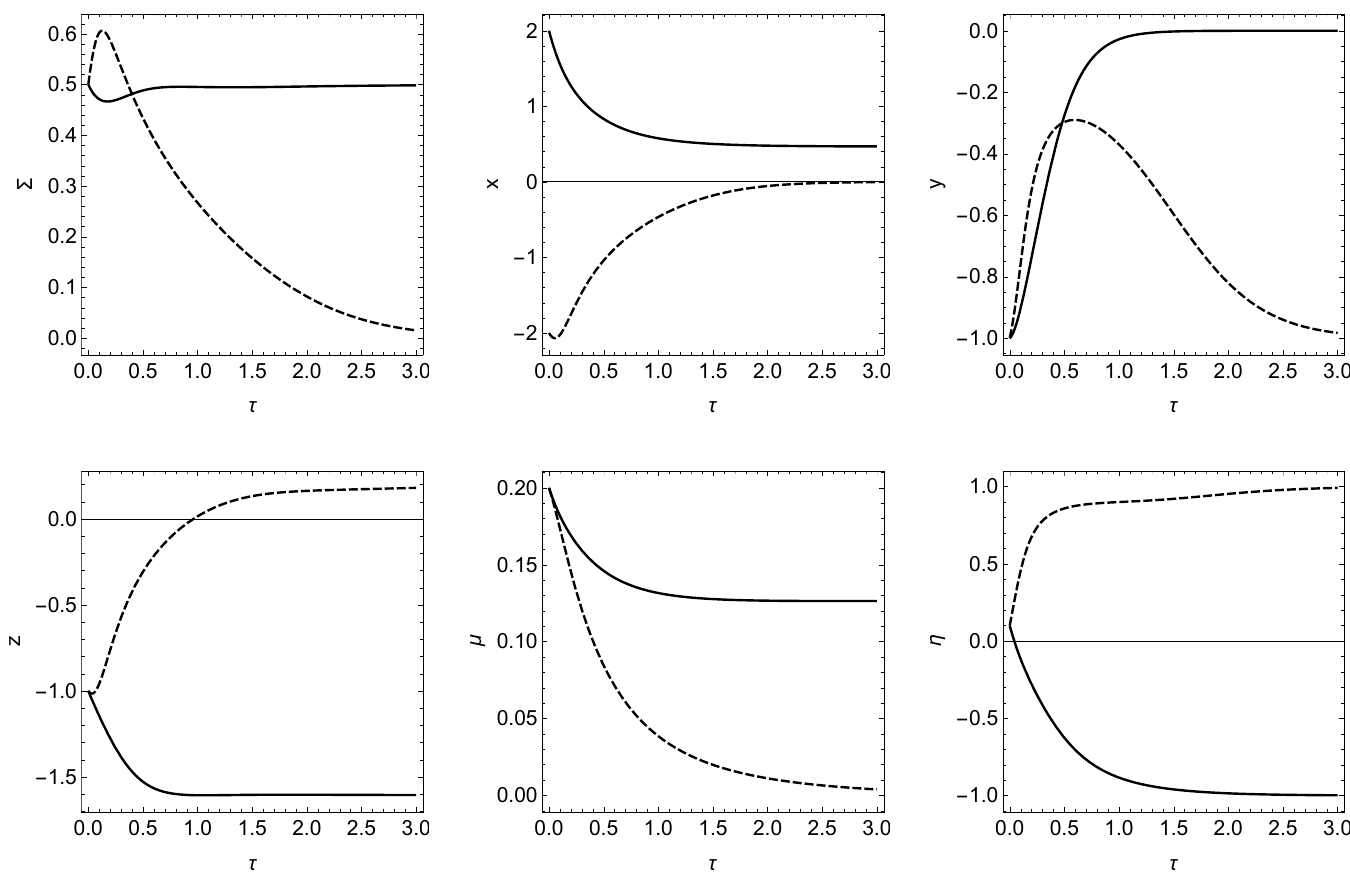}\caption{Qualitative
evolution of the dynamical variables for the system (\ref{dn.01}%
)-(\ref{dn.06}) for $\lambda=1$ and initial conditions $\left(  \Sigma
_{0},y_{0},z_{0},\mu_{0},\eta_{0}\right)  =\left(  \frac{1}{2}%
,-1,-1,0.2,0.1\right)  $. Solid line is for $x_{0}=2$ where the Kasner
solution described by point $B_{3}$ is an attractor and dashed line is for
$x_{0}=-2$ where the de Sitter universe is the attractor.}%
\label{fig3}%
\end{figure}

\begin{figure}[ptb]
\centering\includegraphics[width=1\textwidth]{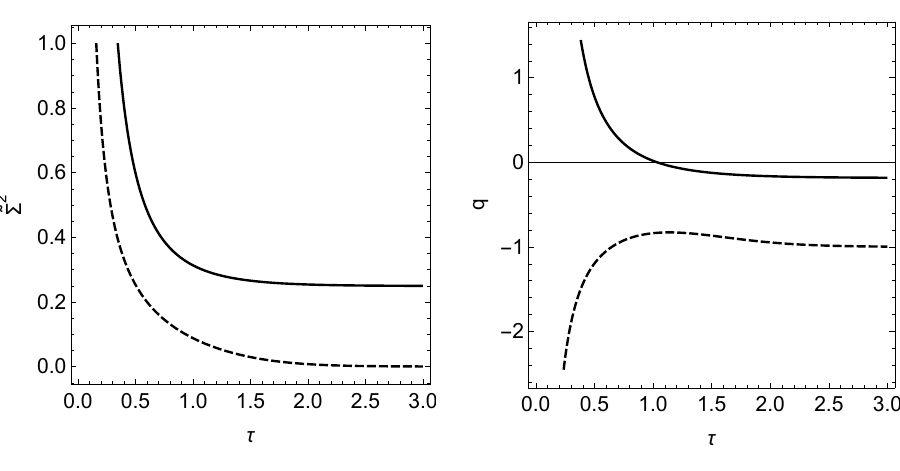}\caption{Qualitative
evolution of the physical variables physical variables $\tilde{\Sigma}^{2}$
and $\tilde{q}$ for the numerical solutions of Fig. \ref{fig3}.}%
\label{fig4}%
\end{figure}

The above results are summarized in Table \ref{tab1}. In Figs. \ref{fig3} and
\ref{fig4} we present the qualitative evolution for the dynamical variables
and for the physical variables $\tilde{\Sigma}^{2}$ and $\tilde{q}$ for two
different sets of initial conditions where points $B_{3}$ and $F_{4}$ are attractors.

%TCIMACRO{\TeXButton{B}{\begin{table}[tbp] \centering}}%
%BeginExpansion
\begin{table}[tbp] \centering
%EndExpansion
\caption{Stationary points and physical parameters.}%
\begin{tabular}
[c]{ccccccc}\hline\hline
\textbf{Point} & \textbf{Spacetime} & \textbf{Tilted} & $\mathbf{\omega}_{m}$
& $\mathbf{\tilde{\Sigma}}^{2}$ & $\mathbf{\tilde{q}}$ & \textbf{Stability}%
\\\hline
$T_{1}$ & Kantowski-Sachs & Yes & $0$ & $\frac{1}{4}$ & $-1+\frac{1}%
{2\sqrt{1+\mu^{2}}}$ & Saddle\\
$T_{2}$ & Kantowski-Sachs & Yes & $0$ & $\frac{1}{4}$ & $-1+\frac{1}%
{2\sqrt{1+\mu^{2}}}$ & Saddle\\
$B_{1}$ & Kasner & No & $0$ & $1$ & $0$ & Saddle\\
$B_{2}$ & Kasner & No & $0$ & $1$ & $0$ & Saddle\\
$B_{3}$ & Kasner & No & $0$ & $1$ & $0$ & Attractor\\
$B_{4}$ & Kasner & No & $0$ & $1$ & $0$ & Saddle\\
$K_{1}$ & Kantowski-Sachs & No & $0$ & $1$ & $-1$ & Saddle\\
$K_{2}$ & Kantowski-Sachs & No & $0$ & $1$ & $-1$ & Saddle\\
$F_{1}$ & FLRW & No & $1$ & $0$ & $0$ & Saddle\\
$F_{2}$ & FLRW & No & $1$ & $0$ & $0$ & Saddle\\
$F_{3}$ & FLRW & No & $0$ & $0$ & $-1$ & Source\\
$F_{4}$ & FLRW & No & $0$ & $0$ & $-1$ & Attractor\\\hline\hline
\end{tabular}
\label{tab1}%
%TCIMACRO{\TeXButton{E}{\end{table}}}%
%BeginExpansion
\end{table}%
%EndExpansion

\subsection{Extreme limit}

We investigate the case of the extreme tilted observer, that is,
$\mu\rightarrow+\infty$. At the extreme tilted, the deceleration parameter
$\tilde{q}$ and the anisotropic parameter $\tilde{\Sigma}^{2}$ are simplified
as $\tilde{q}=-1$ and $\tilde{\Sigma}^{2}=\frac{1}{4}$. Thus at the extreme
tilt scenario the observable universe is always anisotropic and accelerated.

In order to perform this analysis we employ the change of variable $\mu
=\frac{1}{\hat{\mu}}$, and we determine the stationary points for the
dynamical system (\ref{dn.01})-(\ref{dn.06}) with $\hat{\mu}=0$.

The new stationary points $\left(  \Sigma\left(  P\right)  ,x\left(  P\right)
,y\left(  P\right)  ,z\left(  P\right)  ,\hat{\mu}\left(  P\right)
\rightarrow0,\eta\left(  P\right)  \right)  $ are

\begin{itemize}
\item
\begin{align*}
\hat{B}_{1}  &  =\left(  -1,0,0,z_{1},0,-1\right)  ~,~\hat{B}_{2}=\left(
-1,0,0,z_{2},0,1\right)  ~,~\\
\hat{B}_{3}  &  =\left(  1,0,0,z_{3},0,-1\right)  ~,~\hat{B}_{4}=\left(
1,0,0,z_{3},0,1\right)  ~,~
\end{align*}
which correspond to Kasner solutions. The corresponding eigenvalues of the
linearized system for each point are $\hat{B}_{1}:\left\{
-6,-6,-3,0,0,0\right\}  $, $\hat{B}_{2}:\left\{  6,4,2,-1,0,0\right\}
$,~$\hat{B}_{3}:\left\{  -6,-4,-2,1,0,0\right\}  $ and~$\hat{B}_{4}:\left\{
6,6,3,0,0,0\right\}  $. Hence, $\hat{B}_{2}$,~$\hat{B}_{3}$ are always saddle
points and $\hat{B}_{4}$ is a source point. n Figs. \ref{fig1} and \ref{fig2}
we present the two-dimensional phase-space portraits where point $\hat{B}%
_{1}\,\ $is defined. We observe that point $\hat{B}_{1}$ is always a saddle point.

\item
\[
\hat{K}_{1}=\left(  \frac{1}{2},0,-\frac{3}{4},-\frac{2+\lambda}{2\sqrt{3}%
},0,\frac{1}{2}\right)  ~,~\hat{K}_{2}=\left(  -\frac{1}{2},0,-\frac{3}%
{4},\frac{2+\lambda}{2\sqrt{3}},0,-\frac{1}{2}\right)  ~,
\]
correspond to Kantowski-Sachs solutions with a cosmological constant term. The
eigenvalues are~$\hat{K}_{1}:\left\{  -3,-3,-\frac{3}{2},-\frac{3}{2},\frac
{3}{2},\frac{3}{2}\right\}  $ and~$\hat{K}_{2}:\left\{  3,3,\frac{3}{2}%
,\frac{3}{2},-\frac{3}{2},-\frac{3}{2}\right\}  $, from where we infer that
that the asymptotic solutions are always unstable and the $\hat{K}_{1}%
,~\hat{K}_{2}$ are saddle points.

\item
\[
\hat{F}_{3}=\left(  0,0,-1,\frac{\lambda-2}{2\sqrt{3}},0,-1\right)  ~,~\hat
{F}_{4}=\left(  0,0,-1,-\frac{\lambda-2}{2\sqrt{3}},0,1\right)  ,
\]
describe spatially flat FLRW spacetimes dominated by the cosmological
constant. We calculate the eigenvalues $\hat{F}_{3}:\left\{
4,3,3,3,2,-1\right\}  $ and~$\hat{F}_{4}:\left\{  -4,-3,-3,-3,-2,1\right\}  $.
Consequently the stationary points are always saddle points.

\item
\[
\hat{D}_{1}=\left(  2,0,-3,\frac{2+\lambda}{\sqrt{3}},0,-1\right)  ~,~\hat
{D}_{2}=\left(  -2,0,-3,-\frac{2+\lambda}{\sqrt{3}},0,-1\right)
\]
describe Bianchi I spacetimes with a cosmological constant, nonzero matter
component $\omega_{m}\left(  \hat{D}_{1,2}\right)  =-6$ and eigenvalues
$\hat{D}_{1}:\left\{  \frac{3}{2}\left(  1+\sqrt{17}\right)  ,\frac{3}%
{2}\left(  1-\sqrt{17}\right)  ,6,3,3,3\right\}  $, $\hat{D}_{2}:\left\{
-\frac{3}{2}\left(  1+\sqrt{17}\right)  ,-\frac{3}{2}\left(  1-\sqrt
{17}\right)  ,-6,-3,-3,-3\right\}  $. Hence, the points are always saddle points.
\end{itemize}

The above results for the extreme tilted scenario are summarized in Table
\ref{tab2}.

\begin{figure}[ptb]
\centering\includegraphics[width=1\textwidth]{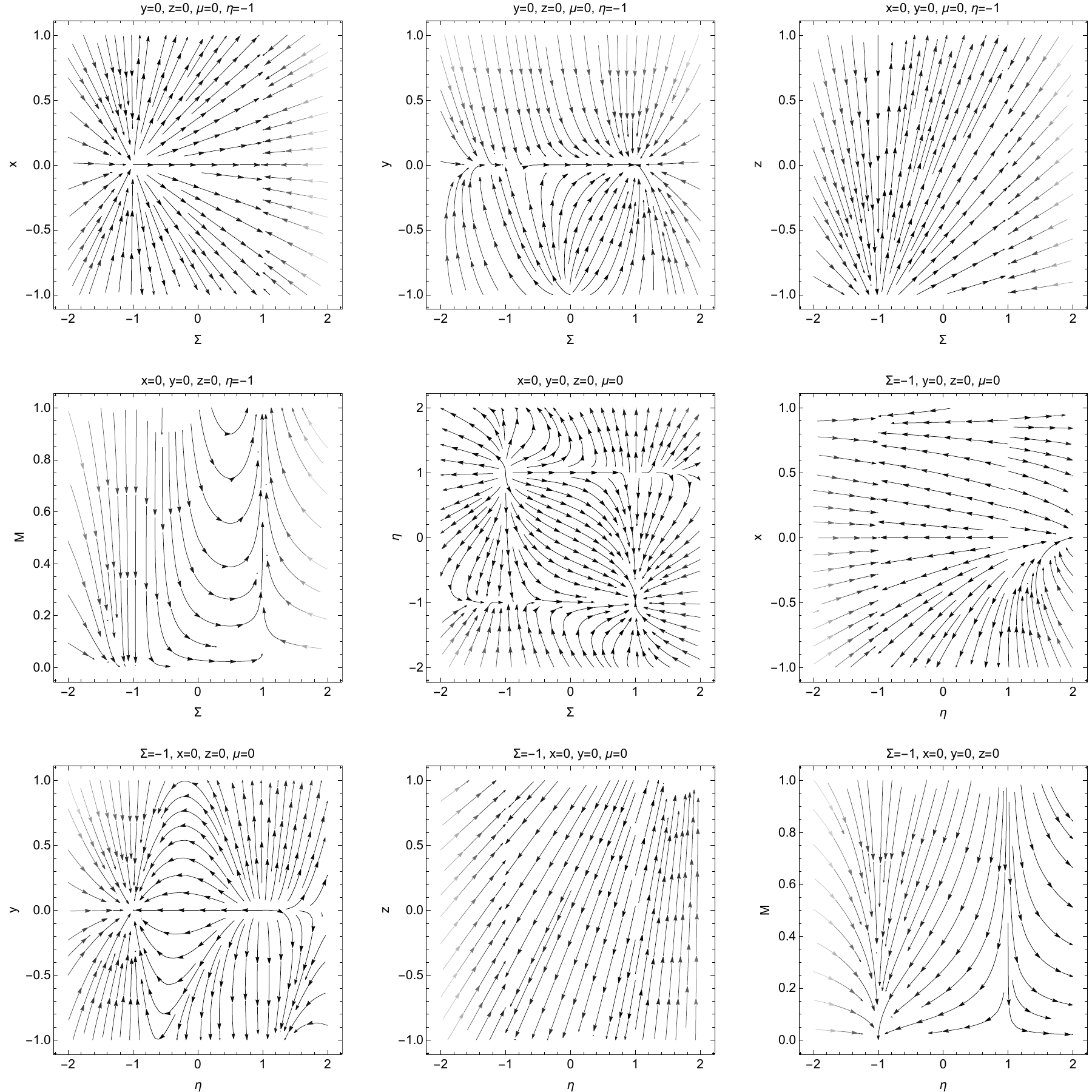}\caption{2D
Phase-space portraits for the dynamical system (\ref{dn.01})-(\ref{dn.06}) on
the surfaces where point $\hat{B}_{1}$ exists. The plots are for $\lambda=1$.
We observe that $\hat{B}_{1}$ is a saddle point.}%
\label{fig5}%
\end{figure}

\begin{figure}[ptb]
\centering\includegraphics[width=1\textwidth]{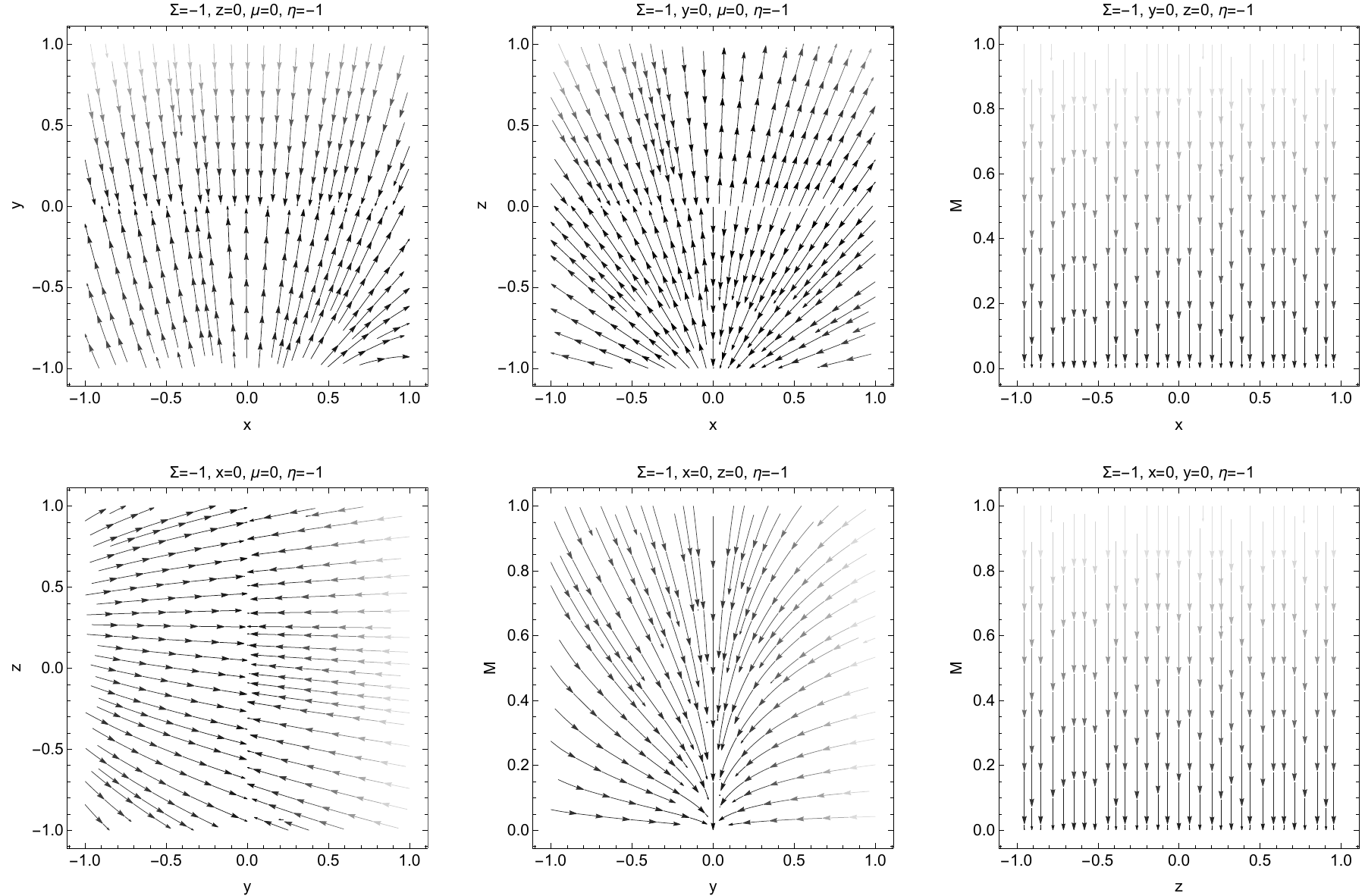}\caption{2D
Phase-space portraits for the dynamical system (\ref{dn.01})-(\ref{dn.06}) on
the surfaces where point $\hat{B}_{1}$ exists. The plots are for $\lambda=1$.
We observe that $\hat{B}_{1}$ is a saddle point.}%
\label{fig6}%
\end{figure}%

%TCIMACRO{\TeXButton{B}{\begin{table}[tbp] \centering}}%
%BeginExpansion
\begin{table}[tbp] \centering
%EndExpansion
\caption{Stationary points and physical parameters at the extreme tilted scenario.}%
\begin{tabular}
[c]{ccccc}\hline\hline
\textbf{Point} & $\mathbf{\omega}_{m}$ & $\mathbf{\tilde{\Sigma}}^{2}$ &
$\mathbf{\tilde{q}}$ & \textbf{Stability}\\
$\hat{B}_{1}$ & $0$ & $\frac{1}{4}$ & $-1$ & Saddle\\
$\hat{B}_{2}$ & $0$ & $\frac{1}{4}$ & $-1$ & Saddle\\
$\hat{B}_{3}$ & $0$ & $\frac{1}{4}$ & $-1$ & Saddle\\
$\hat{B}_{4}$ & $0$ & $\frac{1}{4}$ & $-1$ & Source\\
$\hat{K}_{1}$ & $0$ & $\frac{1}{4}$ & $-1$ & Saddle\\
$\hat{K}_{2}$ & $0$ & $\frac{1}{4}$ & $-1$ & Saddle\\
$\hat{F}_{3}$ & $0$ & $\frac{1}{4}$ & $-1$ & Saddle\\
$\hat{F}_{4}$ & $0$ & $\frac{1}{4}$ & $-1$ & Saddle\\
$\hat{D}_{1}$ & $\neq0$ & $\frac{1}{4}$ & $-1$ & Saddle\\
$\hat{D}_{2}$ & $\neq0$ & $\frac{1}{4}$ & $-1$ & Saddle\\\hline\hline
\end{tabular}
\label{tab2}%
%TCIMACRO{\TeXButton{E}{\end{table}}}%
%BeginExpansion
\end{table}%
%EndExpansion

\section{Conclusions}

\label{sec6}

In General Relativity, and consequently in STEGR, the Einstein tensor for the
Kantowski-Sachs universe is diagonal. This implies that the matter source
allowed in this universe should be described by a diagonal energy-momentum
tensor. In the case of a perfect fluid, the velocity should align with that of
the comoving observer, perpendicular to the three-dimensional spacelike hypersurface.

However, in modified theories like STEGR within a Kantowski-Sachs geometry, it
is possible to consider a tilted velocity for the matter source, resulting in
a dipole universe with a Kantowski-Sachs geometry. To explore this concept,
within the framework of $f(Q)$-gravity, we examined a perfect fluid with a
velocity tilted perpendicular to the two-dimensional spacelike sphere. Our
investigation focused on the phase-space analysis, determining the global
evolution of the physical parameters.

In this model, initial conditions can allow for the fluid to possess a tilted
velocity, implying a violation of the cosmological principle in the early
universe. However, the unique attractors of the model describe the limit of
STEGR with or without a cosmological constant. Specifically, these attractors
correspond to the anisotropic Kasner universe and the isotropic de Sitter
universe. Hence, this dipole gravitational model in $f\left(  Q\right)
$-gravity leads to universe with zero spatially curvature. For initial
conditions where in the asymptotic limit $Q\rightarrow0$, the attractor is the
anisotropic spacetime, while the isotropic de Sitter universe is recovered
when $Q\rightarrow Q_{0},~Q_{0}$ is a nonzero constant. Consequently, we can
infer that the cosmological principle is an attractor of $f\left(  Q\right)  $-gravity.

In the previous analysis we considered the power-law $f\left(  Q\right)
$-theory, which leads to the power-law potential $V\left(  \phi\right)  $ and
the constant parameter $\lambda$. Nevertheless, for arbitrary function
$f\left(  Q\right)  $, that is, scalar field potential $V\left(  \phi\right)
$, parameter $\lambda$ is not a constant. It is evolution it is given the
differential equation%
\begin{equation}
\frac{d\lambda}{d\tau}=\frac{2}{3}\lambda x\left(  \lambda\Gamma\left(
\lambda\right)  +\left(  1-\lambda\right)  \right)  ,~ \label{dla}%
\end{equation}
with $\Gamma\left(  \lambda\right)  =\frac{V_{,\phi\phi}V}{\left(  V_{,\phi
}\right)  ^{2}}$. Hence, the stationary points for the new dynamical system
should solve equation $\lambda x\left(  \lambda\Gamma\left(  \lambda\right)
+\left(  1-\lambda\right)  \right)  =0$. We end with two families of points,
those with $\lambda=\lambda_{0}$ such that $\lambda_{0}\left(  \lambda
_{0}\Gamma\left(  \lambda_{0}\right)  +\left(  1-\lambda_{0}\right)  \right)
=0$, and the points with $x=0$. While the general dynamics and evolution are
affected by a general potential, it is clear that there are not new families
of solutions for arbitrary potential, and the analysis for the power-law
function covers all main families of solutions, as discussed for the FLRW case
in \cite{an01}.

While this work specifically considers $f(Q)$-geometry, these results are
generalizable to other modified symmetric teleparallel theories, as well as to
other Bianchi models. It will be of interest to investigate the existence of
spherical symmetric solutions with this kind of matter source.

\begin{acknowledgments}
The author thanks Prof. Sibu Moyo, Prof. Nikolaos Dimakis, the Stellenbosch
University and the Universidad de La Frontera for the hospitality provided
while part of this work was carried out. AP thanks the support of VRIDT
through Resoluci\'{o}n VRIDT No. 096/2022, Resoluci\'{o}n VRIDT No. 098/2022.
\end{acknowledgments}


\begin{thebibliography}{99}                                                                                               %


\bibitem {t1}L.\ Verde, T.\ Treu and A.G. Riess, Nature Astronomy 3, 891 (2019)

\bibitem {t2}E. Abdalla et al., JHEAstroph. 34, 49 (2022)

\bibitem {t3}L. Perivolaropoulos and F. Skara,\ New Astronomy Reviews 95,
101659 (2022)

\bibitem {t4}E. Di Valentino, O. Mena, S. Pan, L. Visinelli, W. Yang, A.
Melchiorri, D. F. Mota, A. G. Riess and J. Silk, Class. Quantum Grav. 38,
153001 (2021)

\bibitem {in1}A. Krasi\'{n}ski, \textit{Inhomogeneous Cosmological Models},
Cambridge University Press, New York (2006)

\bibitem {in2}J.D. Barrow and M.P. D\c{a}browski, Phys. Rev. D 55, 630 (1997)

\bibitem {in3}P. Szekeres, Phys. Rev.\ D 12, 2941 (1975)

\bibitem {kras}A. Krasi\'{n}ski, \textit{Inhomogeneous Cosmological Models,
}Cambridge U.P., Cambridge, (1997)

\bibitem {Mis69}\ C.W. Misner, Astroph. J. 151, 431 (1968)

\bibitem {jacobs2}K.C. Jacobs, Astrophys J. 153, 661 (1968)\ 

\bibitem {collins}C.B Collins and S.W. Hawking, Astroph. J. 180, 317 (1973)

\bibitem {JB1}J.D. Barrow, Mon. Not. R. astron. Soc. 175, 359 (1976)

\bibitem {JB2}J.D. Barrow and D.H. Sonoda, Phys. Reports, 139, 1 (1986)

\bibitem {jb33}L.G. Jensen and J.A. Stein-Schabes, Phys.\ Rev.\ D 34, 931 (1986)

\bibitem {f3}M.S. Turner and L.M. Widrow, Phys. Rev. Lett. 57, 2237 (1986)

\bibitem {pa1}A.A. Abolhasani, R. Emami and H. Firouzhahi, JCAP 05, 016 (2014)

\bibitem {pa2}E. Dimastrogiovanni, M. Fasiello and L. Pinol, JCAP 09, 031 (2022)

\bibitem {pl1}M. Plionis, MNRAS 234, 401 (1988)

\bibitem {pl2}F. Sorrenti, R. Durrer and M. Kunz, JCAP 11, 054 (2023)

\bibitem {pl3}C. Bonvin, R. Durrer and M. Kunz, Phys. Rev. Lett. 96, 191302 (2006)

\bibitem {pl4}T. Nadolny, R. Durrer, M. Kunz and H. Padmanabhan, JCAP 11, 009 (2021)

\bibitem {val1}E. Di Valentino, A.\ Melchiorri and J. Silk, Nature Astronomy
4, 196 (2020)

\bibitem {ke1}A.R. King and G.F.R. Ellis, Commun. math. Phys. 31. 209 (1973)

\bibitem {ke2}C.B. Collins and G.F.R. Ellis, Phys.\ Reports 56, 65 (1979)

\bibitem {hew}C.G. Hewitt and J. Wainwright, Phys. Rev. D 46, 4242 (1992)

\bibitem {col0}S. Hervik, R. van den Hoogen and A. Coley, Class. Quantum Grav.
22, 607 (2005)

\bibitem {col0a}A.A. Coley and S. Hervik, Class. Quantum Grav. 22, 579 (2005)

\bibitem {col0b}A.A. Coley, S. Hervik and W.C. Lim, Class. Quantum Grav. 23,
3573 (2006)

\bibitem {col1}A.A.\ Coley and D.J. Mc Manus, Phys. Rev. D 54, 6095 (1996)

\bibitem {pl2018}Planck Collaboration: Y. Akrami et al., A\&A 641, A7 (2020)

\bibitem {st1}C.G. Tsagas, Eur. Phys. J. C 82, 521 (2022)

\bibitem {st2}K. Asvesta, L. Kazantizidis, L. Perivolaropoulos and
C.G.\ Tsagas, MNRAS 513, 2394 (2022)

\bibitem {tt1}C. Krishnan, R. Mondol and M.M. Sheikh-Jabbari, JCAP 07, 020 (2023)

\bibitem {tt2}E. Ebrahimian, C. Krishnan, R. Mondol and M.M. Sheikh-Jabbari,
Towards A Realistic Dipole Cosmology: The Dipole $\Lambda$CDM Model (2023) [arXiv:2305.16177]

\bibitem {Nester:1998mp}M. Hohmann, Phys. Rev. D 104, 124077 (2021)

\bibitem {Koivisto2}J. B. Jim\'{e}nez, L. Heisenberg and T. S. Koivisto, Phys.
Rev. D 98, 044048 (2018)

\bibitem {Koivisto3}J. B. Jim\'{e}nez, L. Heisenberg, T. S. Koivisto and S.
Pekar, Phys. Rev. D 101, 103507 (2020)

\bibitem {gg1}V. Gakis, M. Kr\v{s}\v{s}\'{a}k, J.L. Said and E.N. Saridakis,
Phys. Rev.\ D 101, 064024 (2020)

\bibitem {jjd1}L. J\"{a}rv and L. Pati, Phys. Rev. D 109, 064069 (2024)

\bibitem {pal2}N.\ Dimakis, K.J. Duffy, A. Giacomini, A. Yu. Kamenshchik, G.
Leon and A. Paliathanasis, Phys. Dark Univ. 44, 101436 (2024)

\bibitem {Eisenhart}L. P. Eisenhart, Non-Riemannian Geometry, American
Mathematical Society, Colloquium Publications Vol. VIII, New York, (1927)

\bibitem {revh}L. Heisenberg, Physics Reports 1066, 1 (2024)

\bibitem {ww0}L. Atayde and N. Frusciante, Phys. Rev. D 104, 064052 (2021)

\bibitem {ww1}R. Solanki, A. De and P. K. Sahoo, Phys. Dark Universe 36,
100996 (2022)

\bibitem {ww2}F. K. Anagnostopoulos, S. Basilakos and E. N. Saridakis, Phys.
Lett. B 822, 136634 (2021)

\bibitem {ww3}N. Dimakis, A. Paliathanasis and T. Christodoulakis, Class.
Quant. Grav. 38, 225003 (2021)

\bibitem {ww5}W. Khyllep, A. Paliathanasis and J. Dutta,\ Phys. Rev. D 103,
103521 (2021)

\bibitem {ww6}A. Lymperis, JCAP 11, 018 (2022)

\bibitem {ww8}H. Shabani, A. De and T.-H. Loo, Eur. Phys. J. C 83, 535\ (2023)

\bibitem {ww10}A. Paliathanasis, Phys. Dark Univ. 42, 101355 (2023)

\bibitem {ww11}J. Shi, Eur. Phys. J. C 83, 951 (2023)

\bibitem {ww13}J. Ferreira, T. Barreiro, J.P. Mimoso and N.J. Nunes,
Phys.\ Rev. D 108, 063521 (2023)

\bibitem {ww14}S.A. Narawade, L. Pati, B. Mishra and S.K. Tripathy, Phys. Dark
Univ. 36, 101020 (2022)

\bibitem {col}C. B. Collins, J. Math. Phys. 18, 2116 (1977)

\bibitem {col10}A.A. Shaikh and D. Chakraborty, J. Geom. Phys. 160, 103970 (2021)

\bibitem {sw1}W. Wang, H. Chen and T. Katsuragawa, Phys.\ Rev. D 105, 024060 (2022)

\bibitem {sw2}R.-H. Lin and X.-H. Zhai, Phys. Rev. D 103, 124001 (2021)

\bibitem {sw3}F. D' Ambrosio, S. D. B. Fell, L. Heisenberg and S. Kuhn, Phys.
Rev. D 105, 024042 (2022)

\bibitem {sw4}P.\ Bhar, Fortsch. Phys. 71, 2300074 (2023)

\bibitem {sw5}Z. Hassan, S.\ Ghosh, P.K. Sahoo and V.S.H.\ Rao, Gen.\ Rel.
Grav. 55, 90 (2023)

\bibitem {ppr1}D.A. Gomes, J.B. Jimenez, A.J.\ Cano and T.S. Koivisto,
Phys.\ Rev. Lett. 132, 141401 (2024)

\bibitem {ppr2}L. Heisenberg and M. Hohmann, JCAP 03, 063 (2024)

\bibitem {stg1}Y.M. Hu, Y.\ Zhao, X.\ Ren, B. Wang, E.N. Saridakis and Y.-F.
Cai, JCAP 07, 060 (2023)

\bibitem {Hohmann}M. Hohmann, Phys. Rev. D 104 124077 (2021)

\bibitem {Heis2}F. D' Ambrosio, L. Heisenberg and S. Kuhn, Class. Quantum
Grav. 39 025013 (2022)

\bibitem {Zhao}D. Zhao, Eur. Phys. J. C 82, 303 (2022)

\bibitem {an01}A. Paliathanasis, Phys. Dark\ Univ. 41, 101255 (2023)

\bibitem {ks1}N. Dimakis, M. Roumeliotis, A. Paliathanasis, T.
Christodoulakis, Eur. Phys. J. C 83, 794 (2023)

\bibitem {ks2}A. Milano, K. Dialektopoulos, N. Dimakis, A. Giacomini, H.
Shababi, A. Halder and A. Paliathanasis, Kantowski-Sachs and Bianchi III
dynamics in f(Q)-gravity (2024) [2403.06922]

\bibitem {bb1}F. Esposito, S. Carloni and S. Vignolo, Class. Quantum Grav. 39,
235014 (2022)

\bibitem {bb2}A. De, S. Mandal, J.T. Beh, T.-H. Loo and P.K. Sahoo, Eur. Phys.
J. C 82, 72 (2022)

\bibitem {min1}A. Paliathanasis, N. Dimakis and T. Christodoulakis, Phys. Dark
Univ. 43, 101410 (2024)

\bibitem {ks1p}R. Kantowski and R.K.\ Sachs, J. Math. Phys. 7, 443 (1966)

\bibitem {WE}J. Wainwright and G. F. R. Ellis, Dynamical Systems in Cosmology,
Cambridge University Press (1997)

\bibitem {wa1}E. Weber, J. Math. Phys. 26, 1308 (1985)

\bibitem {wa2}H. Baofa, Int. J. Theor. Phys. 30, 1121 (1991)

\bibitem {wa3}E. Weber, J. Math. Phys. 27, 1578 (1986)

\bibitem {wa5}D.-W. Chiou, Phys. Rev D. 78, 044019 (2008)

\bibitem {wa4}B.C. Xanthopoulos,J. Math. Phys. 33, 1415 (1992)

\bibitem {wa7}L.E. Mendes and A.B. Henriques, Phys. Lett. B 254, 44 (1991)

\bibitem {wa8}S. Byland and D. Scialom, Phys. Rev. D 57, 6065 (1998)
\end{thebibliography}
\end{document}